\documentstyle[prl,twocolumn,aps,epsf]{revtex}
\begin{document}

\title { Multi-pion Bose-Einstein correlations in high energy heavy-ion collisions}
\author{Q.H. Zhang}
\address{Institut f\"ur Theoretische Physik, Universit\"at Regensburg,
D-93040 Regensburg, Germany}
\vfill
\maketitle

\begin{abstract}
 Multi-pion correlations and wavepacket 
size effects on the pion multiplicity distribution, pion 
momentum distribution and two-pion interferometry are studied. It is 
shown that multi-pion Bose-Einstein correlations and the wavepacket size cause 
an abundance of pions at low momentum, increase the mean pion 
multiplicity and decrease both the apparent radius of the source and 
the coherent source parameter derived from two-pion interferometry.  

\end{abstract}

PACS number(s): 25.75 Gz,11.38 Mh, 11.30 Rd.

		\section{Introduction}
The main goal of colliding heavy nuclei at high energy is to produce 
quark-gluon plasma (QGP).
 One of the key quantities related to QGP is 
the energy density.  To determine the energy density, we need to 
know the space-time information of the source which is 
the main objective of Hanbury-Brown-Twiss 
correlations\cite{HBT,GGLP}.  
Two-pion Bose-Einstein(BE) correlations are widely used in high energy 
heavy-ion collisions to provide information on the space-time
structure, degree of coherence and dynamics of the region where 
the pions were produced\cite{He96,BGJ,GKW,APW,ZL}. 

In principle, the 
extension of two-pion interferometry to multi-pion interferometry is straightforward
\cite{G85,WC84,Zajc87,Pratt93,SB92,PGG90,Cramer,CGZ95,ZCG95,Z97,Knox,HZ96,ZC97,MP97,SH,W98}.
Experimentally, ultrarelativistic hadronic and nuclear collisions provide 
the environment for creating dozens, and in some cases hundreds, 
of pions
\cite{NA35,WA80,E802,NA44}. Therefore, it is necessary to consider 
 multi-pion correlations in these 
processes. The bosonic nature of the pion should 
affect the single and two-pion spectra and distort the two-pion 
correlation function. 
There is a kind of 
coherence length which corresponds to the wave packet length of the 
emitter, which cause pions to concentrate at 
low momenta\cite{ZCG96,W97,He971}.  Thus, it is very interesting to analyse the 
effects of the multi-pion 
correlations and wavepacket length  on these variables.

The arrangement of this paper is as follows: The derivation 
of the multi-pion correlation function for a 
totally chaotic source is given in section 2. In section 3, the effects 
of multi-pion correlations on the multiplicity 
distribution are discussed. In section 4, the effects of multi-pion correlations 
on two-pion correlation function are analysed for n-pion events and for 
events with all possible 
multiplicities. In section 5, our results for a 
 source distribution are given. 
Conclusions are given in section 6.

\section{Pion correlation function for a chaotic source}

The general definition of the {\it pure} $n$ pion correlation function 
$C_{n}(\vec{p}_{1},\cdot \cdot \cdot \vec{p}_{n})$ is
\begin{eqnarray}
C_{n}(\vec{p}_{1},\cdot \cdot \cdot \vec{p}_{n})
=\frac{P_{n}(\vec {p}_{1},\cdot \cdot \cdot \vec{p}_{n})}
{\prod_{i=1}^n P_{1}(\vec{p_{i}})}   ,
\end{eqnarray}
where $P_{n}(\vec{p}_{1},\cdot \cdot \cdot \vec{p}_{n})$ is the 
probability of observing $n$-pions with momenta $\{ \vec{p}_{i} \}$
all in the {\it same $n$-pion  event}. 

A state created by a classical pion source is described by
\cite{He96,BGJ,GKW,CGZ94,Pratt84}
\begin{equation}
|\phi>=\exp(i\int d \vec{p} \int d^{4}x j(x) \exp(ipx) c^{+}(\vec{p})
|0>,
\end{equation}
where $c^{+}(\vec {p})$ is the pion-creation operator. $j(x)$ is the 
pion current, which can be expressed as 
\begin{equation}
j(x)=\int d^{4}x' d^{4}p j(x',p) \nu(x') \exp(-ip(x-x')) .
\end{equation}
Here $j(x',p)$ is the probability amplitude of finding a pion 
with momentum $p$ , emitted by the emitter at $x'$. $\nu(x')=\exp(i\phi(x'))$ 
is a random phase factor which has been extracted from $j(x',p)$.  
All emitters are uncorrelated in coordinate space when assuming
\begin{equation}
\{\nu^{*}(x')\nu(x)\}=\delta^{4}(x'-x)   .
\end{equation}
Here $\{\cdot\cdot\cdot\}$ means phase average which is defined as\cite{He96}
\begin{eqnarray}
\{\prod_{i=1}^{n}\nu^*(x_i)\prod_{j=1}^{m}\nu(y_j)\}
&\sim& \int_0^{2\pi} 
\prod_{i=1}^{n}\prod_{j=1}^{m}\frac{d\phi(x_i)}{2\pi}\frac{d\phi(y_j)}{2\pi}
\nonumber\\
&&\nu^*(x_i)\nu(y_j).
\end{eqnarray}
Eq.(4) is the ideal case. 
In more realistic cases, each chaotic emitter 
has a small coherent wave packet length, and 
Eq.(4) can be replaced by 
\begin{eqnarray}
\{\nu^{*}(x')\nu(x)\}&=&\frac{1}{\delta^{4}}
\exp[-\frac{(x_{1}-x'_{1})^{2}}{\delta^{2}}
-\frac{(x_{2}-x'_{2})^{2}}{\delta^{2}}
\nonumber\\
&&
-\frac{(x_{3}-x'_{3})^{2}}{\delta^{2}}
-\frac{(x_{0}-x'_{0})^{2}}{\delta^{2}}]  .
\end{eqnarray}
Here $\delta $ is a parameter which determines the coherent length (time) 
scale of the emitter{\footnote{ $\delta$ is smaller than the total 
size ($ R$)  of the source which consists of all emitters}}. 
For simplicity, the same 
coherent scale is taken 
for both space and time at the moment.  
The above formula shows that two emitters within the range of 
$\delta$ can be seen as one emitter, while two-emitters out of this
range are incoherent.  From Eq.(5) we have 
\begin{equation}
\{\prod_{i=1}^{n}\nu^{*}(x_i)\}=\{\prod_{j=1}^{m}\nu(x_j)\}=0.    
\end{equation}

The coherent state can be expanded in Fock-Space as 
\begin{eqnarray}
|\phi>=\sum_{n=0}^{\infty}
\frac{(i \int  j(x) e^{ipx} c^{+}(p) d\vec{p} d^{4}x)^{n}}{n!}|0>
=\sum_{n=0}^{\infty}|n>,
\end{eqnarray}
with
\begin{equation}
|n>=\frac{(i \int  j(x) e^{ipx} c^{+}(p) d\vec{p} dx)^{n}}{n!}|0>.
\end{equation}
Here $|n>$ is an $n$-pion state. According to the definition 
of Eq.(1), the {\it pure} $n$-pion correlation function can be re-expressed as
{\footnote{I would like to express my thanks to the referee and Dr. Hans Feldmeier who make this 
definition more clear.}}
\begin{eqnarray}
C_{n}(\vec{p}_{1},\cdot \cdot \cdot \vec{p}_{n})
&&=
\nonumber\\
&&\frac{\{<n|c^{+}(\vec{p}_{1})\cdot \cdot \cdot c^{+}(\vec{p}_{n})c(\vec{p}_{n})
\cdot\cdot\cdot c(\vec{p}_{1})|n>\}}
{\prod_{i=1}^{n}\{<1|c^{+}(\vec{p}_{i})c(\vec{p}_{i})|1>\}}   .
\end{eqnarray}

From above definition, the {\it pure} two-pion correlation function 
 reads (see Appendix)
\begin{eqnarray}
C_{2}(\vec{p}_{1},\vec{p}_{2})&=&
\frac{\{<2|c^{+}(\vec{p}_{1})c^{+}(\vec{p}_{2})c(\vec{p}_{2})c(\vec{p}_{1})
|2>\}}
{\{<1|c^{+}(\vec{p}_{1})c(\vec{p}_{1})|1>\}
\{<1|c^{+}(\vec{p}_{2})c(\vec{p}_{2}|1>\}}
\nonumber\\
&=&1+
\nonumber\\
&&\frac{\{<1|c^{+}(\vec{p}_{1})c(\vec{p}_{2})|1>\}
\{<1|c^{+}(\vec{p}_{2})c(\vec{p}_{1})|1>\}}
{\{<1|c^{+}(\vec{p}_{1})c(\vec{p}_{1})|1>\}
\{<1|c^{+}(\vec{p}_{2})c(\vec{p}_{2})|1>\}}.
\end{eqnarray}

Using the relationship
\begin{equation}
c(\vec{p})|n>=i\int d^{4}x j(x) \exp(ipx) |n-1>
\end{equation}
we have
\begin{eqnarray}
\{<1|c^{+}(\vec{p}_{1})c(\vec{p}_{2})|1>\}&=&
\{\int d^{4}x_{1} d^{4}x_{2} j^{*}(x_{1}) j(x_{2}) 
\nonumber\\
&&\exp(-i(p_{1}x_{1}-p_{2}x_{2}))\}
\nonumber\\
&=&\{\int d^{4}y d^{4}Y j^{*}(Y+y/2)
\nonumber\\
&&
j(Y-y/2)
\exp(-iky -iqY)\}
\nonumber\\
&=&\int d^{4}Y g_{w}(Y,k)\exp(-iq\cdot Y).
\end{eqnarray}

Here $Y=\frac{x_{1}+x_{2}}{2}$ and $y=x_{1}-x_{2}$ are four dimensional  
coordinates, while $ k=\frac{p_{1}+p_{2}}{2}$ and $ q=p_{1}-p_{2}$ are the 
corresponding four dimensional momenta. 
$g_{w}(Y,k)$ is the Wigner 
function , which can be understood as the probability of finding a pion 
at $Y$ with momentum $k$.  It is defined as 
\begin{equation}
g_{w}(Y,k)=\{\int d^{4}y j^{*}(Y+y/2)j(Y-y/2)\exp(-ik\cdot y)\}.
\end{equation}

Inserting Eq.(3) into the above equation we have
\begin{eqnarray}
g_{w}(Y,k)&=&\{\int d^{4}y \exp(-ik\cdot y)
\nonumber\\
	&&\int d^{4}x'j^{*}(x',p_{1})d^4p_{1}e^{ip_{1}(Y+y/2-x')}\nu^{*}(x')\\
	&&\int d^{4}x''j(x'',p_{2})d^4p_{2}e^{-ip_{2}(Y-y/2-x'')}\nu(x'') \} ,
\nonumber
\end{eqnarray}
so the {\it pure} single pion inclusive distribution $P_{1}(\vec{p})$ can be
expressed as:
\begin{equation}
\begin{array}{lcl}
P_{1}(\vec{k})&=&\int g_{w}(Y,k) d^{4}Y\\
	&&\\
	&=&\{ \int d^{4}x' d^{4}x'' d^{4}p_{1}d^{4}p_{2} 
j^{*}(x',p_{1})j(x'',p_{2}) \\
&&\\
&&\nu^{*}(x')\nu(x'')\delta^{4}(k-\frac{p_{1}+p_{2}}{2})
\delta^{4}(p_{1}-p_{2})\\
&&\\
&&
e^{ip_{1}(Y-x')}e^{-ip_{2}(Y-x'')} \}\\
&&\\
&=&\{ \int d^{4}x' d^{4}x'' 
j^{*}(x',k)j(x'',k) \\
&&\\
&&\nu^{*}(x')\nu(x'')e^{-ik(x'-x'')} \}  \\
\end{array}
\end{equation}
with $k_{0}=\sqrt{\vec{k}^{2}+m_{\pi}^{2}}$.  
The two-pion correlation function reads
\begin{eqnarray}
C_{2}(\vec{p}_{1},\vec{p}_{2})&=&
1+
\nonumber\\
&&\frac{\int d^{4}x d^{4}x' g_{w}(x,k) g_{w}(x',k)
\exp(iq\cdot (x-x'))}{\int d^{4}x d^{4}x' g_{w}(x,p_{1})g_{w}(x',p_{2})}.
\end{eqnarray}

Similarly, the {\it pure} $n$-pion correlation function can be expressed as
\begin{equation}
C_{n}(p_{1},p_{2},\cdot \cdot \cdot, p_{n})=\sum_{\sigma} \chi_{1,\sigma(1)}
\chi_{2,\sigma(2)}...\chi_{n,\sigma(n)}
\end{equation}
with 
\begin{equation}
\chi_{i,j}=\chi(p_{i},p_{j})=\frac{\int d^{4}x g_w(x, \frac{(p_{i}+p_{j})}{2}) 
e^{i(p_{i}-p_{j})\cdot x}}{\sqrt{\int d^{4}x d^{4}yg_w(x,p_{i})g_w(y,p_{j})}}.
\end{equation}
Here $\sigma(i)$ denotes the $i$th element of a permutation of the sequence
${1,2,3,\cdot \cdot \cdot, n}$, and the sum over $\sigma$ denotes the 
sum over all $n!$ permutations of this sequence.

In the following, we shall often use the expression of the {\it pure} $n$-pion 
momentum probability distribution $P_{n}(\vec p_1,\cdot\cdot\cdot \vec p_n)$ 
given by
\begin{equation}
P_{n}(\vec p_1,\cdot\cdot\cdot,\vec p_n)
=\sum_{\sigma} \rho_{1,\sigma(1)}\rho_{2,\sigma(2)}...\rho_{n,\sigma(n)}
\end{equation}
with
\begin{equation}
\rho_{i,j}=\rho(p_{i},p_{j})=\frac{1}{n_{0}}
\int d^{4}x g_w(x, \frac{(p_{i}+p_{j})}{2}) 
e^{i(p_{i}-p_{j})x}.
\end{equation}

Here $n_{0}$ is the mean multiplicity without BE correlations, defined by 
\begin{equation}
n_{0}=\int g_w(x,\vec{k})d^{4}x d\vec{k},
\end{equation}

Once the source distribution $g_w(x,k)$ is known, the {\it pure} multi-pion correlation 
function can be calculated from Eq.(18).  

\section{ Effects of multi-pion correlations on pion multiplicity 
distribution}

From the above definition of $|n>$ we have
\begin{eqnarray}
\omega(n)=\{<n|n>\}=\frac{n_{0}^{n}}{n!}\int 
P_{n}(p_1,\cdot\cdot\cdot,p_n)\prod_{i=1}^{n} d\vec p_{i}  
\end{eqnarray}
and 
\begin{eqnarray}
\{<\phi|\phi>\}&=&\sum_{k} \{<k|k>\}
\nonumber\\
&=&\sum_{k}\frac{n_{0}^{k}}{k!}\int d\vec p_{1} \cdot \cdot
\cdot d\vec p_{k} P_{k}(p_1,\cdot\cdot\cdot,p_k)
\nonumber\\
&=&\sum_{k}\omega(k).
\end{eqnarray}
Then the pion-multiplicity distribution becomes
\begin{eqnarray}
P(n)=\frac{\{<n|n>\}}{\{<\phi|\phi>\}}=\frac{\omega(n)}{\sum_{k} \omega(k)}.
\end{eqnarray}
For particles without BE correlations, we have 
\begin{equation}
\rho_{i,j}=\delta_{i,j}\rho_{i,i}
\end{equation}
and
\begin{equation}
\int P_{n}(p_1,\cdot\cdot\cdot p_n) \prod_{i} d\vec{p}_{i}=1, 
\end{equation}
therefore 
\begin{eqnarray}
P(n)=\frac{n_0^n}{n!}e^{-n_0}.
\end{eqnarray}
That is  $P(n)$ reduces to the usual Poisson distribution.
If we take into account multi-pion Bose-Einstein correlations, the 
pion multiplicity distribution is not a Poisson distribution anymore (see Fig.1).
From the above equation, we can also obtain the mean multiplicity $<M>$
\begin{equation}
<M>=\sum_{n}n\cdot P(n) ~~~ .
\end{equation}

\section{Effects of multi-pion correlations on 
two-pion correlation function}
\subsection{ Two-pion correlation function for n-pion events}

For $n (n\ge 2) \pi$ events, the two-pion correlation function can be defined as
{\footnote{This definition takes into account the higher order Bose-Einstein correlations.}}
\begin{equation}
C_{2}^{n}(p_1,p_2)=\frac{P_{2}^{n}(\vec p_{1},\vec p_{2})}{P_{1}^{n}
(\vec p_{1})P_{1}^{n}(\vec p_{2})} ,
\end{equation}
where $P_{i}^{n}(\vec p_{1},\cdot\cdot\cdot\vec p_{i})$ is the modified $i$-pion inclusive
distribution in $n$ pion events.  The 
definition of $P_{i}^{n}(\vec p_{1},\cdot
\cdot \cdot,\vec p_{i})$ is{\footnote{It is easily checked that 
$\int d\vec p_1 \cdot\cdot\cdot d\vec p_i P_i^n(\vec p_1,\cdot\cdot\cdot,\vec p_i) =1.$
}}
\begin{equation}
P_{i}^{n}(\vec p_{1},\cdot \cdot \cdot \vec p_{i})=
\frac{\{<n|c^{+}(\vec p_{i})\cdot \cdot \cdot
c^{+}(\vec p_{1})
c(\vec p_{1})\cdot \cdot \cdot c(\vec p_{i})|n>\}}{\{<n|n>\} n\cdot \cdot \cdot 
(n-i+1)}.
\end{equation}

For an event with multiplicity  $n$, the modified two-pion inclusive and 
single-pion inclusive distributions can be
expressed as 
\begin{equation}
P_{2}^{n}(\vec p_{1},\vec p_{2})=\frac{\int \prod_{i=3}^{n} d\vec p_{i}
P_{n}(\vec p_{1}....\vec p_{n})}{\int \prod_{i=1}^{n} d\vec p_{i}
P_{n}(\vec p_{1}....\vec p_{n})}
\end{equation}
and
\begin{equation}
P_{1}^{n}(\vec p_{1})=\frac{\int \prod_{i=2}^{n} d\vec p_{i}
P_{n}(\vec p_{1}....\vec p_{n})}{\int \prod_{i=1}^{n} d\vec p_{i}
P_{n}(\vec p_{1}....\vec p_{n})}
\end{equation}
respectively.

As $n$ increases, the computation of the above intergrals becomes 
more and more complex.   Now we define the function
\begin{eqnarray}
G_{i}(p,q)&=&n_{0}^{i} \int \rho(p,p_{1}) d\vec p_{1} \rho(p_{1},p_{2})
d \vec p_{2}
\nonumber\\
&& \cdot \cdot \cdot \rho(p_{i-2},p_{i-1})d \vec p_{i-1}
\rho(p_{i-1},q).
\end{eqnarray}

From the expression of $P_{n}(\vec p_1,\cdot\cdot\cdot,\vec p_n)$ (Eq. (20)),
 the two-pion inclusive distribution can be expressed as
\begin{eqnarray}
P_{2}^{n}(\vec p,\vec q)&=&\frac{1}{n(n-1)}\frac{1}{\omega(n)}
\sum_{i=2}^{n}
\nonumber\\
&&[\sum_{m=1}^{i-1}G_{m}(p,p)G_{i-m}(q,q)+
\nonumber\\
&&G_{m}(p,q)
\cdot G_{i-m}(q,p)]\omega(n-i)
\end{eqnarray}
with
\begin{eqnarray}
\omega(n)=\frac{n_{0}^{n}}{n!}\int \prod_{k=1}^{n} d\vec p_{k} 
P_{n}(p_1,\cdot\cdot\cdot,p_n)~~~.
\end{eqnarray}
The single-pion distribution is given as
\begin{eqnarray}
P_{1}^{n}(\vec p)=
\frac{1}{n}\frac{1}{\omega(n)}\sum_{i=1}^{n}G_{i}(p,p)\omega(n-i).
\end{eqnarray}

Now the main problem is to extract the expression of $\omega(n)$. 
From Eq.(37) we have
\begin{eqnarray}
\omega(n)=\frac{1}{n}\sum_{i=1}^{n} C(i)\omega(n-i),
\end{eqnarray}
with
\begin{equation}
C(i)=\int d\vec p G_{i}(p,p).
\end{equation}

Using the above method, the two-pion and 
single pion inclusive distributions
can be calculated for n-pion events. 
$G_{i}(p,q)$ and $C(i)$ can be calculated using Monte-Carlo
 or analytical integration.  

\subsection{ Two-pion correlation function for all events}

For a state $|\phi>$ which contains all possible multiplicities,
the modified single pion spectrum distribution can be expressed as{\footnote{
 In calculating 
$P_1^{\phi}$ we mix all the multiplicities while in calculating $P_1^n$ we fix the pion 
multiplicity as $n$.}}
\begin{eqnarray}
P_{1}^{\phi}(\vec{p})&=&\frac{\{<\phi|c^{+}(\vec p) c(\vec p)|\phi>\}}
{\{<\phi|\phi>\}<M>}
\nonumber\\
&=&\frac{\sum_{n}P_{1}^{n}(\vec p)\cdot n \cdot \omega(n)}{\{<\phi|\phi>\}<M>}\\
&=&\frac{\sum_{i}G_{i}(p,p)\sum_{n}\omega(n-i)}{\{<\phi|\phi>\}<M>}=
\frac{\sum_{i}G_{i}(p,p)}{<M>}
\nonumber
\end{eqnarray}
where Eq.(24) is applied.  The two-pion inclusive 
distribution can be expressed as
\begin{eqnarray}
P_{2}^{\phi}(\vec{p},\vec{q})&=&
\frac{\{<\phi|c^{+}(\vec p)c^{+}(\vec q) c(\vec q)
 c(\vec p)|\phi>\}}
{\{<\phi|\phi>\}<M(M-1)>}
\nonumber\\
&=&\frac{\sum_{n}P_{2}^{n}(\vec p,\vec q) \cdot n \cdot (n-1) \omega(n)}
{\{<\phi|\phi>\}<M(M-1)>}
\nonumber\\
&=&\frac{1}
{\{<\phi|\phi>\}<M(M-1)>}
\sum_{n}
\nonumber\\
&&
[\sum_{i=2}^{n}(\sum_{m=1}^{i-1}G_{m}(p,p)G_{i-m}(q,q)+
\nonumber\\
&&G_{m}(p,q)
\cdot G_{i-m}(q,p))\omega(n-i)]
\nonumber\\
&=&\frac{\sum_{i,j}G_{i}(p,p)G_{j}(q,q)+G_{i}(p,q)
\cdot G_{j}(q,p)}{<M(M-1)>}.
\end{eqnarray}

The two-pion correlation function for all multiplicity distribution is
\begin{equation}
C_{2}^{\phi}(\vec{p},\vec{q})=\frac{P_{2}^{\phi}(\vec{p},\vec{q})}
{P_{1}^{\phi}(\vec{q})P_{1}^{\phi}(\vec{p})}.
\end{equation}

\section {Multi-pion correlations for a chaotic source}

In this section, we will give an example to investigate the 
wave packet length and the multi-pion correlations effects on 
pion multiplicity, single pion and two-pion 
distributions.   

We assume that the chaotic emitter amplitude distribution is
\begin{eqnarray}
j(x,k)&=&\sqrt{n_0} \exp(\frac{-x_{1}^{2}-x_{2}^{2}-x_{3}^{2}}{2R_{0}^{2}})
\delta(x_{0})
\nonumber\\
&&
\exp(-\frac{k_{1}^{2}+k_{2}^{2}+k_{3}^{2}}{2\Delta^{2}}).
\end{eqnarray}
Where $R_0$ and $  \Delta $ are the parameters which 
represents the radius of the chaotic source and the 
momentum range of pions respectively.
Bringing  Eq.(6) and Eq.(43) into 
Eqs.(15-17), we can easily get the source distribution function $g_w(x,K)$ 
\begin{eqnarray}
g_w(x,K)&=& n_0 \cdot (\frac{1}{\pi R_G^2})^\frac{3}{2}
\exp(-\frac{\vec{x}^2}{R_G^2})\delta(x_0)
\nonumber\\
&&
(\frac{1}{\pi \Delta_0^2})^\frac{3}{2}
\exp\{-\frac{\vec{k}^{2}}{\Delta_0^2}\},
\end{eqnarray}
the normalized {\it pure } pion single 
particle spectrum distribution
\begin{equation}
P_{1}(p)=
(\frac{1}{\Delta_0^2\pi})^{\frac{3}{2}}
\exp\{-\frac{\vec{p}^{2}}{\Delta_0^2} \},
\end{equation}
and the {\it pure} two-pion correlation function
\begin{equation}
C_{2}(\vec{p}_1,\vec{p}_2)=1+\exp
\{
-\frac{\vec{q}^{2}}{2}\frac{4R_{0}^{4}}{\delta^{2}+4R_{0}^{2}}
\}
\end{equation}
with 
\begin{eqnarray}
R_G^2=R_0^2+\frac{1}{\Delta^2},~~~ \frac{1}{\Delta_0^2}=\frac{1}{\Delta^2}+
\frac{R_0^2\delta^2}{\delta^2+4R_0^2}  ~~~.
\end{eqnarray}
In the above formula, the definition of $\vec{q}$ is 
$\vec{q}=\vec{p}_{1}-\vec{p}_{2}$ . 
Then the derived radius through {\it pure} two-pion interferometry is 
\begin{equation}
R^{2}=R_{0}^{2}\frac{4R_{0}^{2}}{\delta^{2}+4R_{0}^{2}}.
\end{equation}

It is clear that as the coherence length $\delta$ increases, 
the mean momentum of pions ($\Delta_0$) becomes smaller (Eq.(47)). 
  It can be seen 
clearly from Eq.(48) that as the coherence length $\delta$ increases, the 
apparent  radius derived from {\it pure} two-pion interferometry also 
becomes smaller.   

 $g_w(x,\frac{p+q}{2})$ can be expressed as
\begin{eqnarray}
g_w(x,\frac{p+q}{2})&=&n_0 \cdot 
\frac{1}{(\pi R_{G}^{2})^{3/2}}e^{-\frac{r^{2}}{R_{G}^{2}}}  
\nonumber\\
&&
\frac{1}{(\pi \Delta_{0}^{2})^{3/2}}e^{-\frac{(\vec{p}+\vec{q})^2}{4\Delta_{0}^{2}}}
\delta(t) ,
\end{eqnarray}
then 
\begin{eqnarray}
\rho(\vec{p},\vec{q})&=&\frac{1}{n_0} \int g_w(x, \frac{p+q}{2})e^{i(p-q)x}dx
\nonumber\\
&=&\frac{1}{(2\pi \Delta_{0}^{2})^{3/2}}
e^{-\frac{(\vec{p}-\vec{q})^{2}R_{G}^{2}}{4}}e^{-\frac{(\vec{p}+\vec{q})^2}{4\Delta_{0}^{2}}}.
\end{eqnarray}

Defining
\begin{eqnarray}
G_{n}(\vec{p},\vec{q})&=&n_0^n\int \rho(\vec{p},\vec{p}_{1}) 
\prod_{i=1,n-2} d\vec p_{i} 
\rho(\vec{p}_{i},\vec{p}_{i+1})
\nonumber\\
&&
d\vec p_{n-1} 
\rho(\vec{p}_{n-1},\vec{q})
\end{eqnarray}
and using Eq.(50),  we easily obtain 
\begin{equation}
G_{n}(\vec{p},\vec{q})=
n_0^n\cdot \alpha_{n} e^{-a_{n}(\vec{p}^{2}+\vec{q}^{2})+g_{n} \vec p \cdot \vec q},
\end{equation}
where
\begin{equation}
a_{n+1}=a_1-\frac{g_1^2}{4b_{n}},~
b_{n}=a_{n}+ a_1,~
g_{n+1}=\frac{g_{n}g_1}{2b_{n}},
\end{equation}
and
\begin{equation}
\alpha_{n+1}=\alpha_{n}(\frac{1}{\Delta_{0}^{2}})^{3/2}(\frac{1}{b_{n}})^{3/2},
\end{equation}
with 
\begin{equation}
a_{1}=\frac{R_{G}^{2}}{4}+\frac{1}{4\Delta_{0}^{2}}, ~ 
g_{1}=\frac{R_{G}^{2}}{2}-\frac{1}{2\Delta_0^2} , ~ 
\alpha_{1}=\frac{1}{(\pi \Delta_{0}^{2})^{3/2}}.
\end{equation}

Now we can calculate the pion multiplicity,  
single particle and two-particle pion distributions  
 according to the formula given
in the preceding sections.  The pion multiplicity distribution 
is shown in Fig.1.   It is clear that the probability with high pion 
multiplicity is greater than that for a conventional Poisson 
distribution. This is consistent with the nature of
bosons, that is, the emission of bosons encourages the emission of more
bosons.  We also find that as the radius and temperature decrease, 
the effect of BE correlations on the multiplicity distribution 
becomes larger.  This can be explained by the fact 
that as the radius and the 
temperature decrease the quantum effect become larger. 
It is also clear that as the wavepacket $\delta$ increases, the effect of 
BE correlations on the multiplicity distribution becomes larger. It can be 
seen clearly from Eq.(47) that as $\delta$ increases, $\Delta_0$ 
(the effective temperature)
 becomes smaller therefore the quantum effect becomes larger.
\vskip -0.0cm
\begin{figure}[h]\epsfxsize=8cm
\centerline{\epsfbox{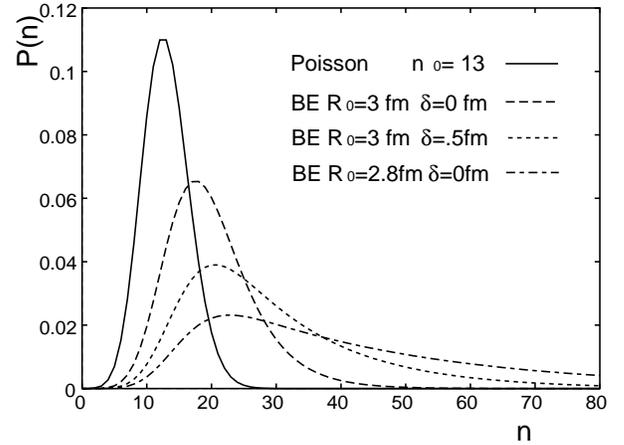}}
\caption{\it 
Pion multiplicity distribution with and 
without (solid cure) Bose-Einstein correlations.
The input value of $\Delta $ and $ n_0$ are
$0.25 GeV$ and $ 13$. 
The dashed, dotted and dot-dashed line corresponds to  
 $R_0= 3fm$ ,$\delta = 0 fm$; 
 $R_0= 3fm$ ,$\delta = 0.5  fm$ 
and  $R_0= 2.8 fm$ ,$\delta = 0 fm$ respectively.}
\end{figure}
In Fig.2, the mean multiplicity $<M>$  is plotted against the parameter $n_0$,
the mean multiplicity without BE correlations. 
When $n_0$ is not very large, $<M>$ increases with increasing 
$n_0$.  However,  at a critical point $n_c$, the mean 
multiplicity $<M>$ becomes to be infinity.  
From Eqs. (39-40), the mean multiplicity can be expressed as:
\begin{eqnarray}
<M>=\sum_i C(i)=\sum_i n_0^i \alpha_i (\frac{\pi}{2a_i-g_i})^{3/2}.  
\end{eqnarray}
It is clear that the above equation will remain convergence provided that 
\begin{eqnarray}
\lim_{n\rightarrow \infty}\frac{C(n+1)}{ C(n)}
&=&\lim_{n\rightarrow \infty}n_0\cdot (\frac{1}{\Delta_0^2})^{\frac{3}{2}}
\nonumber\\
&&
(\frac{1}{b_n})^{3/2}(\frac{2a_{n+1}-g_{n+1}}{2a_n-g_n})^{3/2} < 1.
\end{eqnarray}
From Eqs.(53-55), we have 
\begin{eqnarray}
\lim_{n\rightarrow \infty}a_n&=&\frac{R_G}{2\Delta_0},\nonumber\\
\lim_{n\rightarrow \infty}b_n&=&(\frac{R_G}{2}+\frac{1}{2\Delta_0})^2,
\nonumber\\
\lim_{n\rightarrow \infty}g_n&=&0.
\end{eqnarray}
\vskip -0.0cm
\begin{figure}[h]\epsfxsize=8cm
\centerline{\epsfbox{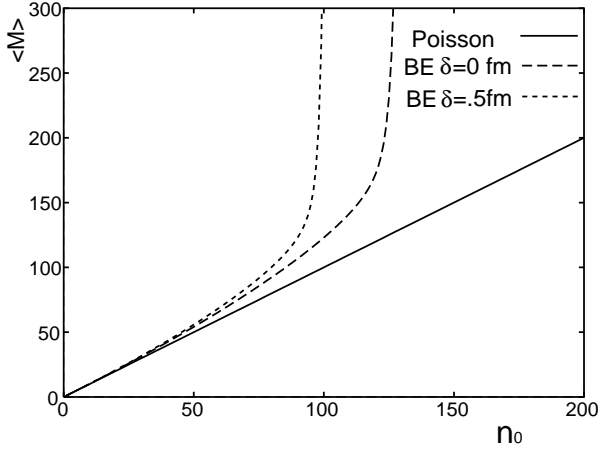}}
\caption{\it 
The mean multiplicity $<M>$ vs. $n_{0}$. The solid line
corresponds to the case without BE correlation $(n_0 = <M>)$.  
The dashed and dotted lines corresponds to $\delta =0 fm$ and 
$\delta =0.5 fm$ respectively.  
The input value 
of $R_0$ and $\Delta$ are $5 fm$ and $0.36 GeV$.}
\end{figure}
Eq.(56) remains convergence if 
\begin{eqnarray}
n_0 &<& \frac{1}{8}(1+R_G \Delta_0)^3
\nonumber\\
&=&
\frac{1}{8}\big(1+\sqrt{1+\frac{4R_0^2\Delta^2}{4+(R_0^2\Delta^2+1)\delta^2/R_0^2}}
\big)^3=n_c.
\end{eqnarray}
We denote this critical pion multiplicity as $n_c$. 
It is very interesting to notice that the critical multiplicity $n_c$ depends 
only on $R_0\Delta_0 $ and $\delta/R_0$. 
The critical multiplicity $n_c$ vs. $R_0\Delta_0 $ for different values of 
$\delta/R_0$ are shown in Fig.3. 
It is clear that as $\delta /R_0 $ increases the corresponding
critical multiplicity becomes smaller. 

\vskip -0.0cm
\begin{figure}[h]\epsfxsize=8cm
\centerline{\epsfbox{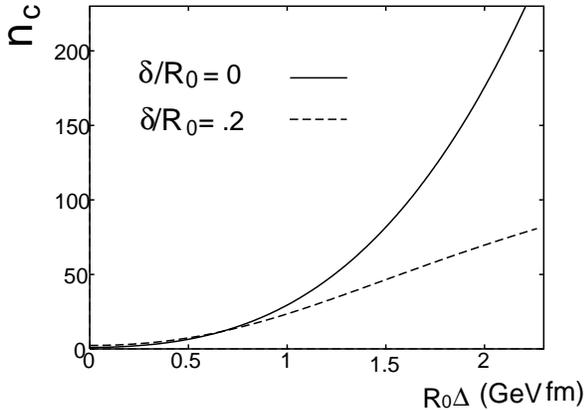}}
\caption{\it 
The critical multiplicity $n_c$ vs. $ R_0\Delta $. The solid line 
and dashed line corresponds to $\delta/R_0 =0 $ and 
$\delta/R_0 =0.2 $ respectively.}
\end{figure}
Including the multipion BE correlations, the single particle momentum 
distribution for events with fixed multiplicity is given in Fig.4.  
It is shown that, as the multiplicity of pion increases, 
the number of pions at low momentum becomes larger.  
This has been observed in many experiments\cite{NAtt}. 
Many different explanations of 
the experimental results have been proposed\cite{PP,KS92}.  
Some of those\cite{KS92} stated that the enhancement of single 
pion spectrum at low momentum spectrum is 
due to the two-pion Bose-Einstein correlations.  
Our results show that the enhancement of single pion spectrum at low 
momentum caused by mult-pion correlations is more pronounced than that 
caused by two-particle correlations.  So a detail analysis of 
the reason for the enhancement of single particle spectrum 
at low momentum is still necessary.  The effects of the wavepacket 
 on the pion spectrum distribution are also 
shown in the Fig.4.. It is clear that the effects of 
the wavepacket on the pion spectrum distribution becomes larger 
for larger pion multiplicity. 
 The single particle momentum distributions  
for total events are shown in Fig.5. Similar features as figure 4 are obtained.

\vskip -0.0cm
\begin{figure}[h]\epsfxsize=8cm
\centerline{\epsfbox{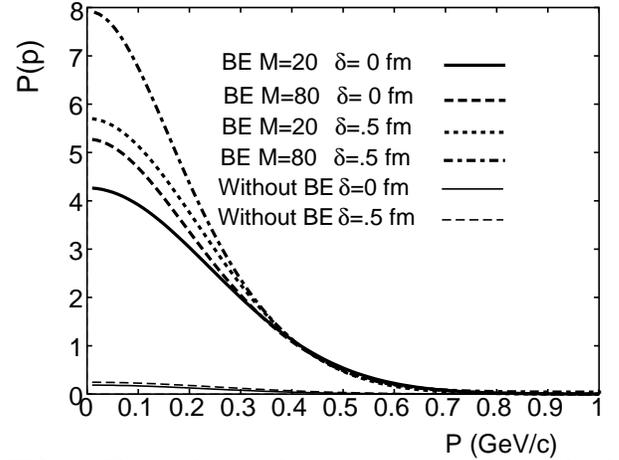}}
\caption{\it 
The single particle momentum distribution for different multiplicities. 
The wider solid line and wider dotted line corresponds to 
multiplicity $M=20$, $\delta =0 fm$ and 
$ M=20 $ ,$\delta =0.5 fm $ respectively.  
The wider dashed line and wider dot-dashed line 
corresponds to multiplicity 
$M=80$, $\delta=0 fm$ and $M=80$, $\delta=0.5 fm$ respectively.  
The thin solid line and the thin dashed line corresponds to 
the input momentum distribution with $\delta=0 fm$ and $\delta=0.5 fm $
respectively. 
The input value 
of $R_0$ and $\Delta$ are $5 fm$ and $0.36 GeV$.}
\end{figure}
The effects of multi-pion correlations on the two-pion correlation 
function are shown in 
Fig.6,  where we fixed the multiplicity of the events.
It can clearly be seen that as the multiplicity of the event increases,  
the two-pion correlation function has a lower chaoticity, though
the actual source is totally chaotic.  
 The apparent 
radius derived from two-pion interferometry 
also becomes smaller. 
Two-pion correlation functions for  different  
 mean multiplicity are shown in Fig.7 .  A similar property to that in Fig.6 can be seen. 
It is clear that the effects of the wavepacket on two-pion correlation  
becomes larger for larger multiplicity.

\vskip -0.0cm
\begin{figure}[h]\epsfxsize=8cm
\centerline{\epsfbox{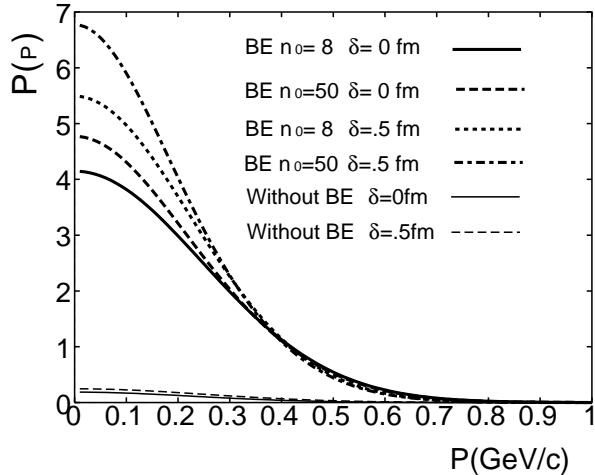}}
\caption{\it
The single particle momentum distribution for different $n_0$. 
The wider solid line and wider dotted line corresponds to 
 $n_0=8$, $\delta =0 fm$ and 
$ n_0=8$ ,$\delta =0.5 fm $ respectively.  
The wider dashed line and wider dot-dashed line 
corresponds to 
$n_0=50$, $\delta=0 fm$ and $n_0=50$, $\delta=0.5 fm$ respectively.  
The thin solid line and thin dashed line corresponds to 
the input momentum distribution with $\delta=0 fm$ and $\delta=0.5 fm$
The input value 
of $R_0$ and $\Delta$ are $5 fm$ and $0.36 GeV$.}
\end{figure}
\vskip -0.0cm
\begin{figure}[h]\epsfxsize=8cm
\centerline{\epsfbox{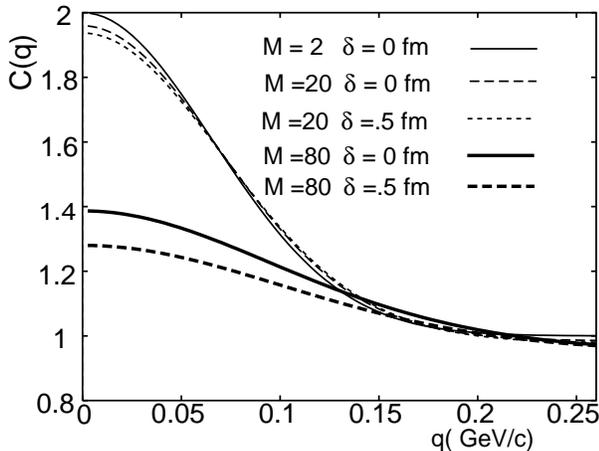}}
\caption{\it 
Two-pion correlation function for fixed multiplicity events.  
The thin solid line correspond to multiplicity $M=2$ and $\delta=0 fm$.
The thin dashed line and thin dotted line corresponds to multiplicity
$M=20$,$\delta=0 fm$ and $M=20$,$\delta=0.5 fm$ respectively.  
The wider solid line and the wider dashed line corresponds to 
$M=80$, $\delta=0 fm$ and $M=80$, $\delta =0.5 fm $ respectively. 
The input value 
of $R_0$ and $\Delta$ are $3 fm$ and $0.36 GeV$.}
\end{figure}
\vskip -0.0cm
\begin{figure}[h]\epsfxsize=8cm
\centerline{\epsfbox{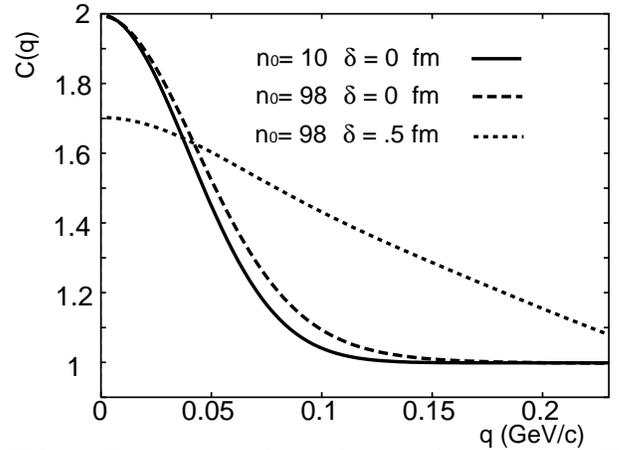}}
\caption{\it 
Two-pion correlation function for all events.  The solid line 
corresponds to $n_0=10, \delta = 0fm$.
The dashed line and the dotted line corresponds to $n_0=98, \delta=0 fm$ 
and $n_0=98 ,\delta = 0.5 fm $ respectively. 
 The input value 
of $R_0$ and $\Delta$ are $5 fm$ and $0.36 GeV$.}
\end{figure}
\section{Conclusions}
Early in the next decade two heavy-ion accelerators, the Relativity 
Heavy Ion Collider(RHIC) and the Large Hadron Collider(LHC), will 
provide the environment for creating hundreds of pions.  
So the multi-pion Bose-Einstein correlations effect in 
those process  should be studied. 
In this paper, the multi-pion Bose-Einstein correlations and wavepacket effects 
on pion multiplicity distribution, pion spectrum distribution and 
two-pion interferometry have been studied.  
 It has been shown that multi-pion 
Bose-Einstein correlations and wavepacket cause an abundance of pions at 
the low momentum, increase the pion's mean multiplicity and decrease both the apparent
radius of the source and the coherent source parameter derived from 
two-pion interferometry.  For larger pion multiplicity,
 the effects of the wavepacket on the 
pion multiplicity distribution, pion spectrum distribution and two-pion interferometry become 
 larger. 

\begin{center}
{\bf Acknowledgement}
\end{center}
The author would like to express his gratitude to Drs. U. Heinz, Y. Pang, W.Q. Chao, 
,C.S. Gao, P. Scotto and Urs. Wiedemann for helpful discussions.  Also the author would like 
to express his thanks to Drs. H. Feldmeier,J. Knoll, D. Mi\'skowiec and V. Toneev for 
helpful discussions and suggestions during the author visit GSI. The author thank Prof. U. Heinz 
for his reading the manuscript.  Finally the author would 
like to express his thanks to the referee for his helful comments and suggestion 
about the paper. This work was supported by the Alexander von Humboldt foundation in 
Germany.

\begin{center}
{\bf Appendix}
\end{center}
The derivation of Eq. (11) will be given in this appendix. From Eq. (10) we 
have 
\begin{equation}
C_2(\vec{p}_1,\vec{p}_2)=\frac{\{<2|c^+(\vec{p}_1)c^+(\vec{p}_2)
c(\vec{p}_2)c(\vec{p}_1)|2>\}}
{\{<1|c^+(\vec{p}_1)c(\vec{p}_1)|1>\}\{<1|c^+(\vec{p}_2)c(\vec{p}_2)|1>\}},
\end{equation}
with
\begin{equation}
|n>=\frac{(i\int j(x) e^{ipx} c^+(\vec{p})d\vec{p}d^4x)^n}
{n!}|0> .
\end{equation}
Using the relationship
\begin{equation}
c(\vec{p})|n>=i\int d^4 x j(x) \exp(ip\cdot x)|n-1>,
\end{equation}
we have 
\begin{eqnarray}
c(\vec{p}_1)c(\vec{p}_2)|2>&=&i\int d^4x j(x) \exp(ip\cdot x)
\nonumber\\
&&
i\int d^4y j(y) \exp(ip\cdot y) |0>.
\end{eqnarray}
Then the two pion spectrum distribution reads 
\begin{eqnarray}
&\{<2|c^+(\vec{p}_1)c^+(\vec{p}_2)c(\vec{p}_2)c(\vec{p}_1)|2>\}&=
\nonumber\\
&\{
\int d^4x j^*(x_1) \exp(ip\cdot x_1)
\int d^4y j^*(y_1) \exp(ip\cdot y_1)& 
\nonumber\\
&
\int d^4x j(x_2) \exp(-ip\cdot x_2)
\int d^4y j(y_2) \exp(-ip\cdot y_2)\}.&
\end{eqnarray}
Here $\{\cdot\cdot\cdot \}$ means the phase average(Eq.(5)).  Using Eq.(3), Eq.(5) and Eq.(6),
we have the following relationship
\begin{equation}
\{j^*(x_1)\cdot\cdot\cdot j^*(x_n)j(y_1)\cdot\cdot \cdot j(y_m)\}=0 ~~~ n\ne m,
\end{equation}
then 
\begin{eqnarray}
\{j^*(x_1)j^*(y_1)j(x_2)j(y_2)\}&=&\{j^*(x_1)j(x_2)\}\{j^*(y_1)j(y_2)\}
+
\nonumber\\
&&\{j^*(x_1)j(y_2)\}\{j^*(y_1)j(x_2)\}.
\end{eqnarray}
Eq.(64) can be re-expressed as
\begin{eqnarray}
&\{<2|c^+(\vec{p}_1)c^+(\vec{p}_2)c(\vec{p}_2)c(\vec{p}_1)|2>\}
&=
\nonumber\\
&
\{<1|c^+(\vec{p}_1)c(\vec{p}_1)|1>\}\{<1|c^+(\vec{p}_2)c(\vec{p}_2)|1>\}+&
\nonumber\\
&\{<1|c^+(\vec{p}_1)c(\vec{p}_2)|1>\}\{<1|c^+(\vec{p}_2)c(\vec{p}_1)|1>\}.&
\end{eqnarray}
Bringing Eq. (67) into Eq. (60) we obtain Eq. (11).

\newpage
\begin{center}
{\bf Figure Captions}
\end{center}

\end {document}